\documentclass[a4paper,leqno]{article}
\usepackage{amsmath}
\usepackage{amsthm}
\usepackage{graphicx}
\usepackage{latexsym}
\usepackage{amssymb}
\usepackage{amsbsy}
\usepackage[]{authblk}

\DeclareSymbolFont{msbm}{U}{msb}{m}{n}
\DeclareMathSymbol{\C}{\mathalpha}{msbm}{'103}
\DeclareMathSymbol{\R}{\mathalpha}{msbm}{'122}
\DeclareMathSymbol{\Z}{\mathalpha}{msbm}{'132}
\DeclareMathSymbol{\N}{\mathalpha}{msbm}{'116}

\newtheorem{remark}{Remark}

\newtheorem{lgrthm}{Algorithm}

\def\be{\begin{equation}}
\def\ee{\end{equation}}
\def\bea{\begin{eqnarray}}
\def\ba{\begin{array}{l}\displaystyle}
\def\eea{\end{eqnarray}}
\def\ea{\end{array}}

\begin{document}
\title{Direct simulation Monte Carlo schemes for Coulomb interactions in plasmas}
\author[1]{Giacomo Dimarco\footnote
{Corresponding author address: Institut de Math\'{e}matiques de
Toulouse, UMR 5219 Universit\'{e} Paul Sabatier, 118, route de
Narbonne 31062 TOULOUSE Cedex, FRANCE. \\
\emph{E-mail}:giacomo.dimarco@math.univ-toulouse.fr,
caflisch@math.ucla.edu, lorenzo.pareschi@unife.it}}
\author[1]{Russell Caflisch}
\author[2]{Lorenzo Pareschi}

\affil[1]{Universit\'{e} de Toulouse; UPS, INSA, UT1, UTM ;

Institut de Math\'{e}matiques de Toulouse ; F-31062 Toulouse,
France.}

\affil[2]{Mathematics Department, University of California,\\ Los
Angeles (UCLA), California 90095-1555, USA. }

\affil[3]{ Mathematics Department and CMCS,\\ University of Ferrara,
Via Machiavelli 35, 44100 Ferrara, Italy.}

\maketitle

\begin{abstract}
We consider the development of Monte Carlo schemes for molecules
with Coulomb interactions. We generalize the classic algorithms of
Bird and Nanbu-Babovsky for rarefied gas dynamics to the Coulomb
case thanks to the approximation introduced by Bobylev and Nanbu
\cite{Bobylev}. Thus, instead of considering the original Boltzmann
collision operator, the schemes are constructed through the use of
an approximated Boltzmann operator. With the above choice larger
time steps are possible in simulations; moreover the expensive
acceptance-rejection procedure for collisions is avoided and every
particle collides. Error analysis and comparisons with the original
Bobylev-Nanbu (BN) scheme are performed. The numerical results show
agreement with the theoretical convergence rate of the approximated
Boltzmann operator and the better performance of Bird-type schemes
with respect to the original scheme.
\end{abstract}


\section{Introduction}

When a gas is far from the thermodynamical equilibrium, the
description of the system through the fluid equation is not
satisfactory and its fundamentals properties depend upon the
interactions of the particles. Collisional phenomena can be
distinguished for long-range interactions and short-range
interactions. Short-range interactions are typically of rarefied
gases and they are described through the Boltzmann equation
\cite{cercignani}, while long-range interactions are normally
encountered in plasmas and modeled through the
Landau-Fokker-Planck equation \cite{bittencourt}. Nowadays,
numerical simulations of plasmas are receiving a great deal of
attention both in research and in industry thanks to the numerous
applications directly connected to these phenomena. In addition,
there exist many practical situations in which the so-called
Coulomb collisions are fundamental for correctly describing the
plasma dynamics as for instance in magnetic fusion devices.

However, while the literature on Monte Carlo schemes for
short-range forces is wide (\cite{Babovsky}, \cite{bird},
\cite{CPmc}, \cite{CPima},\cite{caflisch}, \cite{Nanbu80}) and
many efficient methods have been developed for treating these
problems, on the other hand, how to construct efficient Monte
Carlo numerical methods for long-range interactions like the
Coulomb potential field it is still not clear. The present work is
a contribution in this direction. We observe that recently, some
important results have been achieved by Nanbu and Bobylev in
\cite{Nanbu} and \cite{Bobylev} on this subject. They proposed a
new Monte Carlo numerical method which permits to efficiently
simulate Coulomb collisions. The performance of this scheme has
been studied in details by Caflisch et al in \cite{wang} and
\cite{Candal}, in comparisons with the classical scheme for
simulating Coulomb collisions by Takizuka and Abe \cite{Abe}.
Monte Carlo methods for plasmas have also been developed by Wang
et al \cite{wang1} while an interesting hybrid method has been
realized and numerically tested by Caflisch et al \cite{cohen} for
accelerating the simulation of Coulomb collisions.

In the case of Coulomb interactions each particle interacts
simultaneously with a large number of other particles, thus multiple
collisional phenomena are involved adding difficulties to the
numerical description. However, these multiple interactions can be
seen as a number of simultaneous binary collisions, each of which
gives a small contribution to the relaxation process through a small
angle scattering between particles. \cite{bittencourt}.

The Landau-Fokker-Planck equation is a valid substitute to the
Boltzmann operator when describing this type of systems. More
precisely, the Landau-Fokker-Planck equation can be derived as an
asymptotic form of the Boltzmann equation in the case of a Coulomb
potential field, in which the large angle deflection of a charged
particle in a multiple Coulomb interaction is considered as a series
of consecutive weak binary collisions \cite{Bobylev, Desvilletts}.
However, the classic Boltzmann collision integral is still able to
describe the interactions, but the typical time between two
consecutive collisions prohibits construction of efficient explicit
schemes, since the resulting time step will be in these cases too
small and an excessive use of the computational resources is needed.
Moreover due to the small effect of a single interaction such
detailed modeling is unnecessary.

In the present work, starting from an approximation derived by
Bobylev and Nanbu \cite{Bobylev} for the Boltzmann collision
operator, we derive a general class of direct Monte Carlo methods in
the same spirit of Monte Carlo schemes for rarefied gas dynamic.
Moreover, keeping separated the discretization of the time
derivative and the approximation of the collision operator, we
perform a series of numerical convergence tests for the approximated
collision operator in order to show that the effective convergence
rate coincides with the one hypothesized in \cite{Bobylev}.

The rest of the paper is organized as follow. In Section 2, we
introduce the Boltzmann and Fokker-Planck equations and their
properties. In Section 3, we derive the approximated Boltzmann
operator and the limiting case in which, for small angle scattering,
it converges to an approximated operator for the
Landau-Fokker-Planck equation. Section 4 concerns the construction
of direct simulation Monte Carlo methods. Several test problems
which show the capabilities of the methods, the differences and
analogies with the original Bobylev-Nanbu (BN) scheme, and the
convergence rates are presented in Section 5. Some final
considerations are discussed in Section 6.

\section{The Boltzmann and the Landau-Fokker-Planck equations}

Consider the Boltzmann equation \be \label{eq:boltz}\frac{\partial
f(x,v,t)}{\partial t}+v\cdot\nabla_x f(x,v,t)+\cdot\nabla_v (a(v)
f(x,v,t))=Q(f,f) \ee with the initial condition \be
 f(x,v,t=0)=f_{0}(x,v) ,
\label{eq:B1} \ee where $f=f(x,v,t)$ is a non negative function
describing the time evolution of the distribution of particles which
move with velocity $v \in \R^3 $ at the position $x \in \Omega
\subset \R^{3}$ at time $ t > 0$. The vector $a(v)$ represents the
acceleration due to the force acting on particles such as gravity,
electric field or magnetic field. The bilinear operator $Q(f,f)$
describes the binary collisions between particles and is given by
\be Q(f,f)=\int_{\R^3}\int_{S^2} B\left(|q|,\frac{q\cdot
n}{|q|}\right) [f(v')f(v'_*)-f(v)f(v_*)]dn dv_*\ee where $S^2$ is
the unit sphere in $\R^3$, $q=v-v_*$, $n\in S^2$ the unit normal.
The post collisional velocity are computed by \be
v'=\frac{1}{2}(v+v_*+|q|n), \ v'_*=\frac{1}{2}(v+v_*-|q|n)  \ee The
collision kernel $B(|q|,q\cdot n/|q|)$, which characterizes the
detail of the interaction, is defined as \be
B(|q|,\cos\theta)=|q|\sigma(|q|,\theta), \ (0 \leq \theta \leq \pi)
\ee Here $\cos \theta=q\cdot n/|q|$ and $\sigma(q,\theta)$ is the
collision cross section at the scattering angle $\theta$, that
correspond to the number of particles scattered per unit time, per
unit of incident flux and per unit of solid angle. We introduce also
the total scattering cross section and the momentum scattering cross
section that will be used in the remainder of the paper \be
\sigma_{tot}(|q|)=2\pi\int_{0}^{\pi}\sigma(|q|,\theta)\sin \theta
d\theta\ee \be
\sigma_{m}(|q|)=2\pi\int_{0}^{\pi}\sigma(|q|,\theta)\sin
\theta(1-\cos \theta)d\theta\ee In the case of hard sphere molecules
the cross section and the collision kernel takes the form \be
\sigma(q,\theta)=\frac{d^2}{4}, \
B(|q|,\theta)=\frac{d^2}{4}|v-v_*|\ee while in the variable hard
sphere case we have \be \sigma(q,\theta)=C_\alpha|v-v_*|^{\alpha-1},
\ B(|q|,\theta)=C_\alpha|v-v_*|^{\alpha}\ee with $C_\alpha$ and
$\alpha$ positive constants. The case $\alpha=0$ is referred as
Maxwellian gas while for $\alpha=1$ we recover the hard sphere
model. In the case of Coulomb interactions the Rutherford formula
holds \be \sigma(|q|,\theta)=\frac{b_0^2}{4 \sin^4(\theta/2)}\ee
where $b_0=e^{2}/(4\pi\epsilon_0 m_r |v-v_*|^2)$, with $e$ the
charge of the particle, $\epsilon_0$ the vacuum permittivity and
$m_r$ the reduced mass, which corresponds to $m/2$, if the particles
are of the same species, with $m$ equal to the mass. Observe that
the above formula implies that the scattering cross section tends to
infinity as the angle $\theta$ tends to zero. In order to obtain
finite and meaningful values for the total and the momentum cross
section it is necessary to introduce a cut-off value for the impact
parameter. The cut-off value is justified by the shielding effect
phenomena, leading to the following values for the total cross
section and the momentum cross section \be
\sigma_{tot}(|q|)=\pi\lambda_d^2\ee \be \sigma_{m}(|q|)=4\pi
b_0^2\log\Lambda\ee with $\lambda_d=(\frac{\epsilon_0 k T}{n
e^2})^{1/2}$ the Debye length and
$\Lambda=\frac{1}{\sin(\theta^{min}/2)}$.

In the case of grazing collisions it is possible to derive from the
Boltzmann operator the Landau-Fokker-Planck operator (see
\cite{Desvilletts} for details) \be
Q^L(f,f)=\frac{1}{8}\frac{\partial}{\partial
v_i}\int_{\R^3}|q|\sigma_{m}(|q|)((|q|^2)\delta_{ij}-q_iq_j)\times
\left(\frac{\partial}{\partial v_j}-\frac{\partial}{\partial
v_{*j}}\right)f(v)f(v_*)dv_*\label{eq:landau}\ee In the next section
we will see how it is possible to construct numerical schemes
starting from the Boltzmann equation which approximate the Landau
operator (\ref{eq:landau}).

\section{The approximated Boltzmann equation}

From now on, we will focus on the space homogeneous equation without
force fields. Once the collision term is solved, the solution of the
full Boltzmann equation can be recovered by computing the transport
and the force term through a time splitting.

Although the divergence of the collision integral has been solved
with the cut-off of the scattering cross section, the simulation of
the Boltzmann equation for Coulomb interactions still represent a
significant challenge, due to the too high computational cost which
is necessary to directly simulate the equations with time explicit
schemes. In fact rewriting Eq. (\ref{eq:boltz}) in the space
homogenous case pointing out the gain and loss term \be
\frac{\partial f}{\partial t}=Q^{+}(f,f)-f(v)L [f](v), \
L[f](v)=\sigma_{tot}(|q|)\int_{\R^3}|q|f(v_*)dv_*\ee \be
Q^+(f,f)=\int_{\R^3}\int_{S^2} B\left(|q|,\frac{q\cdot
n}{|q|}\right) f(v')f(v'_*)dn dv_*\ee it is easy to observe that the
large value of the total collision cross section forces the time
step to be small, thus too many steps become necessary to compute
the final solution, yielding this scheme useless. In fact
discretizing the time derivative we obtain \be f(v,t+\Delta
t)=\Delta t Q^+(f,f)+f(v,t)\left(1-\Delta t
\sigma_{tot}(|q|)\int_{\R^3}|q|f(v_*)dv_*\right)\ee now if we want
to preserve a probabilistic interpretation we need the coefficients
to be positive, thus $\Delta t$ has to be extremely small if
$\sigma_{tot}(|q|)$ is very large.

Recently an approximated Boltzmann operator has been developed by
Bobylev and Nanbu (\cite{Bobylev}), which permits use of larger time
steps during the simulation even in the case of Coulomb collisions.
Here we try to generalize this approach in order to construct Direct
Monte Carlo schemes for small particles interactions.

Rewrite equation (\ref{eq:boltz}) in the homogenous case in the
following form \be \label{eq:boltz1}\frac{\partial f}{\partial
t}=\int_{\R^3} JF(U,q)dv_* \ee where $U=(v+v_*)/2$ denotes the
center of mass velocity, and \be F(U,q)\equiv
f(U+q/2)f(U-q/2)=f(v)f(v_*)\ee while the operator $J$ is defined as
\be JF(U,|q|\omega)=\int_{S^2}B(|q|,\omega\cdot
n)[F(U,|q|n)-F(U,|q|\omega)]dn\ee
 with $\omega=q/|q|$. If we approximate the operator $J$ in equation (\ref{eq:boltz1}) by \be J=\frac{1}{\tau}(\exp(\tau
J)-\widehat{I})\ee where $\widehat{I}$ is the identity operator and
$\tau$ are assumed to be small, the equation reads \be
\frac{\partial f}{\partial t}=\frac{1}{\tau}\int_{\R^3}(\exp(\tau
J)-\widehat{I})F(U,q)dv_*=\frac{1}{\tau}(P_{\tau}(f,f)-\varrho
f)\label{eq:Boltz2}\ee with \be P_{\tau}(f,f)=\int_{\R^3}\exp(\tau
J)f(v)f(v_*)dv_*\ee The operator $\exp(\tau J)$ can be written as
\be \label{eq:exp}\exp(\tau J)\psi(\omega)=\int_{S^2}B_{\tau}(\omega
\cdot n,|q|)\psi(n)dn \ee where $\psi(\omega)$ is an arbitrary
function and \be \label{eq:green} B_{\tau}(\omega\cdot
n,|q|)=\sum_{l=0}^{\infty}\frac{2l+1}{4
\pi}\exp(-\lambda_{l}(|q|)\tau)P_{l}(\omega\cdot n)\ee is the Green
function, with $P_l(\omega\cdot n)$ the Legendre polynomial and
$\lambda_l(|q|)$ equal to \be \lambda_l(|q|)=2 \pi
\int_{-1}^{1}B(\mu,|q|)(1-P_l(\mu))d\mu \ee where $\mu=\omega\cdot
n, \ -1\leq\mu\leq 1$ Using the above expression we obtain \be
P_{\tau}(f,f)=\int_{\R^3\times S^2}B_{\tau}(|q|,\frac{q\cdot
n}{|q|})f(v')f(v'_*)dn dv_*\ee

Note that \be \int_{S^2}B_{\tau}(\omega \cdot n,|q|)=1 \ee

\subsection{A first order approximation for the Landau-Fokker-Planck
equation} \label{sec:FLP} Assume now that the scattering cross
section $\sigma(|q|,\theta)$ is concentrated at small angle near
$\theta\approx 0$, thus $B_{\tau}(|q|,\mu)$ is concentrated near
$\mu=1$. In that situation it is possible to derive the following
formal approximation \be \lambda_l(u) \simeq
2\pi\int_{-1}^{1}B_{\tau}(|q|,\mu)(1-P_l(1)+(1-\mu)P_l'(1))d\mu=\pi
l (l+1)\int_{-1}^{1}B_{\tau}(|q|,\mu)(1-\mu)d\mu\ee where
$P_l'(1)=l(l+1)/2$. The approximate Green function reads \be
B_{\tau}(\mu,|q|)\simeq
B_{\tau}^L(\mu,|q|)=\sum_{l=0}^{\infty}\frac{2l+1}{4\pi}P_l(\mu)\exp\left(-\frac{l(l+1)}{2}|q|\sigma_{m}(|q|)\tau\right)\label{eq:Green}\ee
The superscript $L$ in equation (\ref{eq:Green}) means that equation
(\ref{eq:Boltz2}) with the above kernel approximates the
Landau-Fokker-Planck equation. For a formal proof we refer to the
paper of Nanbu and Bobylev (\cite{Bobylev}).

Consider now the case of a Coulomb potential field in a single
component gas or plasma. This choice, with the cut-off of the
scattering angle introduced in the previous section, leads to the
following approximated equation of order $0(\tau)$ \be
\frac{\partial f}{\partial t}=\frac{1}{\tau}\left(\int_{\R^3\times
S^2}D\left(\frac{q\cdot
n}{|q|},\frac{\tau}{2\varrho\tau_1}\right)f(v',t)f(v_*',t)dn
dv_*-\varrho f(v,t)\right)\ee where \be
\frac{1}{\tau_1}=4\pi\left(\frac{e^2}{4\pi\epsilon_0 m_r
}\right)^{2}\frac{\varrho \ln \Lambda}{|q|^3} \ee and \be
D(\mu,\tau_0)=\sum_{l=0}^{\infty}\frac{2l+1}{4\pi}P_l(\mu)\exp(-l(l+1)\tau_0)
. \label{eq:Bobylev}\ee

\section{DSMC schemes for Coulomb Interactions}

Note that is not necessary to work with the collisional kernel
$D(\mu,\tau_0)$ computed above, instead a simpler function
$D^*(\mu,\tau_0)$ can be used, preserving the same accuracy
$0(\tau)$, if the following condition remain satisfied \be
D^*(\mu,\tau_0)\geq 0, \ 2\pi\int_{-1}^{1}D^*(\mu,\tau_0)d\mu=1\ee
\be \lim_{\tau_0\rightarrow
0}D^*(\mu,\tau_0)=\frac{1}{2\pi}\delta(1-\mu)\ee \be
\lim_{\tau_0\rightarrow
0}\frac{2\pi}{\tau_0}\int_{-1}^{1}[D^*(\mu,\tau_0)-D(\mu,\tau_0)]P_l(\mu)d\mu=0
. \ee One possible substitution is represented by \be
D^*(\mu,\tau_0)=\frac{A}{4\pi \sinh A}\exp(\mu A) \ee where
$A=A(\tau)$ satisfy \be \coth
A-\frac{1}{A}=\exp^{-\frac{\tau}{\varrho\tau_1}}\ee It is now clear
that it is possible to apply slightly modified versions of the
standard direct Monte Carlo algorithms for Maxwell molecules to the
equation \be \frac{\partial f}{\partial
t}=\frac{1}{\tau}(P_{\tau}^*(f,f)-\varrho f)\label{eq:timedisc}\ee
with \be P_{\tau}^*(f,f)=\int_{\R^3\times S^2}D^*\left(\frac{q\cdot
n}{|q|},\frac{\tau}{2\varrho\tau_1}\right)f(v',t)f(v_*',t)dn dv_*
\ee The only difference is the way the angle is sampled. In most of
the DSMC methods (Hard sphere or Variable Hard Sphere scattering
models) the angle is sampled uniformly over the sphere, while here
is sampled accordingly to $D^*(\mu,\tau_0)$. Let us discretize the
time and denote $f^n(v)$ the approximation of $f(v,n\Delta t)$, the
forward Euler scheme can be used to solve Eq. (\ref{eq:timedisc})
\be f^{n+1}=\left(1-\frac{\varrho\Delta
t}{\tau}\right)f^{n}+\frac{\varrho \Delta
t}{\tau}P^{*}_{\tau}(f,f)\label{eq:dsmc}\ee This equation has the
following probabilistic interpretation: a particle with velocity
$v_i$ will not collide with probability $\left(1-\varrho\Delta
t/\tau\right)$ and it will collide with probability $\varrho\Delta
t/\tau$ accordingly to the collision law described by
$P^*_{\tau}(f,f)$. Observe that the probabilistic interpretation
holds till $\varrho\Delta t \leq \tau$, otherwise the coefficient in
front of $f^n$ becomes negative. Note that taking the limit of the
above relation, $\varrho\Delta t = \tau $, leads to the scheme of
Nanbu and Bobylev. The possibility to take different values of
$\Delta t\leq \tau/\varrho$ permits reduction of the statistical
fluctuations and reduction of the error due to the time
discretization at no additional cost since, in contrast to the
Variable Hard sphere case, here no acceptance-rejection procedure is
present.

Hence a Monte Carlo algorithm for the solution of the approximated
space homogeneous Landau-Fokker-Planck equations reads as follows

\begin{lgrthm}[Nanbu-Babovsky (NB) for Coulomb Interactions]{~}
\label{al:MC1}
\begin{enumerate}
\item Given $N$ samples $v_k^0$ with $k=1,2,..,N$ computed from the initial
distribution function $f(v,t=0)$
\item DO $n=1$ to $n_{TOT}$ with $n_{TOT}=t_{final}/\Delta t$

Given $\{v_{k}^{n},k=1,...,N\}$
\begin{enumerate}
\item Set $N_c=round(\varrho N \Delta t/2\tau)$, where the $round$ is
statistical
\item Select $N_c$ pairs $(i,j)$ uniformly among all possible pairs
\item Perform the collision between $i$ and $j$ particles according
to the following collision law \begin{enumerate}
\item Compute the cumulative scattering angle $\cos \theta$ as \be \label{eq:cos} \cos \theta=\frac{1}{A}\ln(\exp^{-A}+2U \sinh A) \ee
where $U$ is a random number and $A=A(\tau)$ is computed through the
solution of the non linear equation \be \coth
A-\frac{1}{A}=\exp^{-\frac{\tau}{\varrho\tau_1}}\ee
\item With the above value of $\cos \theta$ perform the collision between $i$ and $j$ and compute the post collisional velocity
according to \be v_i'=v_i-\frac{1}{2}(q(1-\cos \theta)+h\sin
\theta)\ee \be v_j'=v_j+\frac{1}{2}(q(1-\cos \theta)+h\sin
\theta)\ee where $q=v_i-v_j$, while h is defined as \be \nonumber
h_x=q_\bot \cos \epsilon \ee \be \nonumber h_y=-(q_y q_x \cos
\epsilon + q q_z \sin \epsilon)/q_\bot \ee \be \nonumber h_z=-(q_z
q_x \cos \epsilon - q q_y \sin \epsilon)/q_\bot \ee where
$q_\bot=(q_y^2+q_z^2)^{1/2}$ and $\epsilon=2\pi U_1$ with $U_1$ a
random number
\item set $v^{n+1}_i=v^{'}_i$ and $v^{n+1}_j=v^{'}_j$
\end{enumerate}
\item Set $v^{n+1}_i=v_i$ for the particles that
have not been selected
\end{enumerate}
END DO
\end{enumerate}
\end{lgrthm}

We noticed the structure of the approximate operator analyzed in the
previous section is similar to the structure of the classical
Boltzmann collision integral. Thus it is possible to construct, in
the same spirit of Maxwell molecules for rarefied gas dynamic, a
Monte Carlo scheme based on the classical Bird method.

From the inspection of the approximated Landau operator, it follows
that the average number of significant collisions in a time step
$\Delta t$ is given by \be N_c=\frac{N}{2}\frac{\varrho \Delta
t}{\tau}\ee which means that the average time between two collisions
is given by \be \frac{\Delta t}{N_c}=\frac{2\tau}{\varrho N}\ee The
Bird method, in the case of Maxwell molecules, can be seen as a NB
scheme in which the smallest possible time step $\Delta t_1=\Delta
t/ N_c$ is used, in fact only one pair collide each $\Delta t_1$ \be
f^{n+1}=\left(1-\frac{\varrho\Delta
t_1}{\tau}\right)f^{n}+\frac{\varrho \Delta
t_1}{\tau}P^{*}_{\tau}(f,f)=\left(1-\frac{2}{N}\right)f^{n}+\frac{2}{N}P^{*}_{\tau}(f,f)\label{eq:bird}\ee
Hence the Bird algorithm for the approximated Landau-Fokker-Planck
equation reads
\begin{lgrthm}[Bird for Coulomb Interactions]{~} \label{al:MC2}
\begin{enumerate}
\item Given $N$ samples $v_k^0$ with $k=1,2,..,N$ computed from the initial
distribution function $f(v,t=0)$
\item set time counter $t_c=0$
\item set $\Delta t_c=2\tau/\varrho N$
\item DO $n=1$ to $n_{TOT}$ with $n_{TOT}=t_{final}/\Delta t$
\begin{enumerate}
\item repeat
\begin{enumerate}
  \item Select a random pair $(i,j)$ uniformly within all possible
  pairs
  \item perform the collision accordingly to the collision law
  defined in the first algorithm and produce $v^{'}_i$ and $v^{'}_j$
  \item Set $\tilde{v}_i=v^{'}_i$ and $\tilde{v}_j=v^{'}_j$
  \item update the time counter $tc=tc-\Delta t_c$ until $t_c\geq (n+1)\Delta t$
\end{enumerate}
\item Set $v^{n+1}_i=\tilde{v}_i, i=1,...,N$
\end{enumerate}
END DO
\end{enumerate}
\end{lgrthm}

The main difference with respect to the previous algorithm is that
multiple collisions between particles are allowed in $\Delta t$.
Moreover in this case the stability condition can be violated and
$\tau$ can be greater than $\varrho \Delta t$. Thus, as the number
of samples increase to infinity, the Nanbu-Babovsky scheme converge
in probability to the discretized approximated Boltzmann
equation. On the other hand, 
the Bird scheme converges to the solution of the approximated
Boltzmann equation increasing the number of samples, in fact $\Delta
t_1$ approaches zero. Thus for $\tau\rightarrow 0$ the first
converges to the solution of the discretized Landau-Fokker-Planck
equation while the second converges to the exact solution of the
Landau-Fokker-Planck equation.
\begin{remark}
Observe that $\tau$, although in the schemes has a role similar to
the Knudsen number in rarefied gas dynamics, has not a clear
physical meaning. Mathematically the parameter in front of the
collision operator is a measure of the goodness of the
approximation. Once it is fixed it gives a model for the
interactions between particles which approximates the Landau
operator as $\tau$ goes to zero.
\end{remark}

\section{Numerical Tests}

\subsection{Test case}

The behavior of the DSMC schemes is illustrated through a series of
tests in which the relaxation of the velocity distribution function
with anisotropic temperature $T$ is considered. Thus the initial
distribution is taken to be ellipsoidal with $T_x\neq T_y=T_z$. The
initial values for temperature and density are set to
           \be T=4\times Ry, \ \varrho=0.5 \ee
where $Ry$ is the Rydberg constant. The initial difference in the
temperature is fixed to $\Delta T_{0}=0.8$.  The approximate
analytic solution of the Fokker-Planck equation, in the case of
small temperature difference, for $\Delta T=T_x-T_y$ is given by
\cite{Trubnikov} \be \Delta T=\Delta T_0\exp^{-\frac{8}{5\sqrt
{2\pi}}\frac{t}{\tau_T}}\label{eq:Trub}\ee the relaxation time
$\tau_{T}$ corresponds to \be \frac{1}{\tau_{T}}=\frac{\varrho
e^4\log\Lambda}{\pi\sqrt{2}\epsilon_0^2m^{1/2}(kT)^{3/2}}\ee where
the Coulomb logarithm value is fixed to $\log\Lambda=0.5$. The
simulations are run for most of the relaxation process; because all
the schemes reach the same final equilibrium state and start from
the same initial data, our interest is to analyze the different
behaviors of the methods when particles are both far from these two
situations. Thus, fixing $t_{f}=40$ with the values above reported
we obtain $\Delta T_f/\Delta T_0\simeq 0.2$ for the analytic
solution, in the rest of the equilibrium process the schemes and the
analytic solution become closer till they coincide. We remark that
although analytic, the solution is still obtained through
approximations and valid for ranges in which $\Delta T_0$ is small.
This is made more clear by the Figures at the end of the section,
even for very small $\tau$ the schemes do not relax at the same rate
of the approximate formula (\ref{eq:Trub}).

\subsection{Simulations}

Our aim is to perform comparisons between the Bird scheme and the
Bobylev-Nanbu (BN) scheme. The curves which describe the behavior of
the Nanbu-Babovsky (NB) scheme with different choices of the time
step lie between this two extreme cases. In the sequel we will
analyze
\begin{itemize}
  \item the deterministic error;
  \item the statistical fluctuations of the two schemes.
\end{itemize}
The sources of errors for the methods are due to
\begin{itemize}
  \item approximation of the Boltzmann operator;
  \item finite number of particles;
  \item discretization of the time derivative if present;
  \item conservative algorithm used for collisions.
\end{itemize}
In order to make a fair comparison of the methods and to stress the
capacity to describe the relaxation phenomena we try to eliminate
the common sources of errors. First we compute the deterministic
error due to the substitution of the original collision operator
with its approximation and to the discretization of the time
derivative. To that aim we increase the number of samples and
average the solution of M independent realizations removing
statistical fluctuations\be
\overline{u}(t)=\frac{1}{M}\sum_{i=1}^{M}u_i(t) \ee where
$\overline{u}(t)$ indicates the average solution at time $t$ and \be
u_i(t)=\frac{T_i(t)_x-T_i(t)_y}{T(0)_x-T(0)_y}\ee with $i$ the
realization number. The number of samples used in the convergence
analysis test for each realization is N=$2\times 10^6$ while the
number of realizations is $M=5$.   
Observe however that, using the Bird method together with a large
number of particles for each realization, leads to a very accurate
discretization of the time derivative (the effective time step is a
function of 1/N), while with the Babovsky-Nanbu scheme the increase
of the samples number does not affect the treatment of the time
derivative. Note that since no acceptance-rejection procedure is
necessary the two methods have approximately the same computational
cost.
Summarizing the first test tries to measure the deterministic error
computing the numerical order of accuracy with respect to $\tau$ of
the two methods. From the theoretical analysis we expect both
methods to be first order in $\tau$. Note however that for BN method
the error is due to both the approximation of the operator and the
to discretization of the time derivative. The order of accuracy $r$
in $\tau$ is computed as \be r(\tau)=\log_2{R(\tau)}, \
R(\tau)=\frac{|\overline{u}(4\tau)-\overline{u}(2\tau)|}{|\overline{u}(2\tau)-\overline{u}(\tau)|}\label{eq:error}\ee
with $R$ the error ratio. Our second purpose is to measure the
stochastic fluctuations of the two methods. To this aim we compare
the two variances defined as \be
\Sigma^2(\tau,N)=\frac{1}{M}\sum_{i=1}^{M}(u_i-\overline{u})^{2}\ee
Fixing the parameter $\tau$, i.e. the approximation of the Boltzmann
operator, the variances of the two methods are compared for
increasing number of samples starting from $N=100$ to $N=3200$. In
this test the number of realizations is chosen equal to $M=1500$.
Thanks to multiple collisions we expect the Bird scheme to have
slightly less fluctuations with respect to the BN scheme.

\subsection{Results}

\begin{figure}
\begin{center}
\includegraphics[scale=0.58]{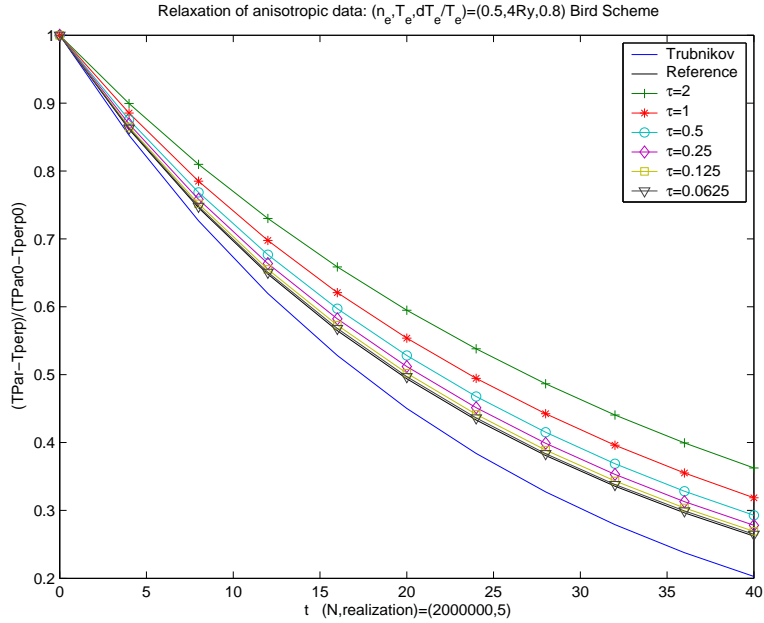}
\caption{Relaxation of the velocity distribution function with
anisotropic initial data for different values of $\tau$. Bird
scheme.}\label{fig:DSMC1}
\end{center}
\end{figure}

\begin{figure}
\begin{center}
\includegraphics[scale=0.58]{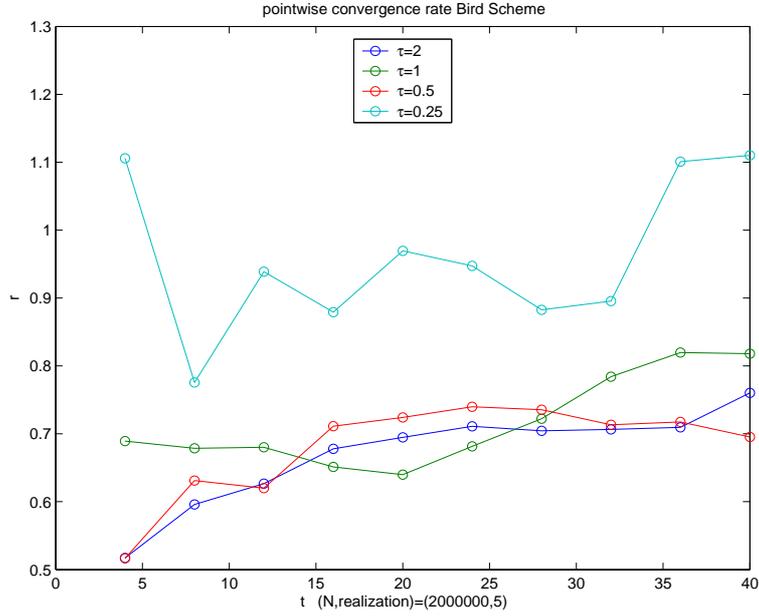}
\caption{Pointwise order of accuracy $r(\tau)=\log2(R(\tau))$ for
the Bird scheme for decreasing values of the parameter
$\tau$.}\label{fig:DSMC2}
\end{center}
\end{figure}

\begin{figure}
\begin{center}
\includegraphics[scale=0.58]{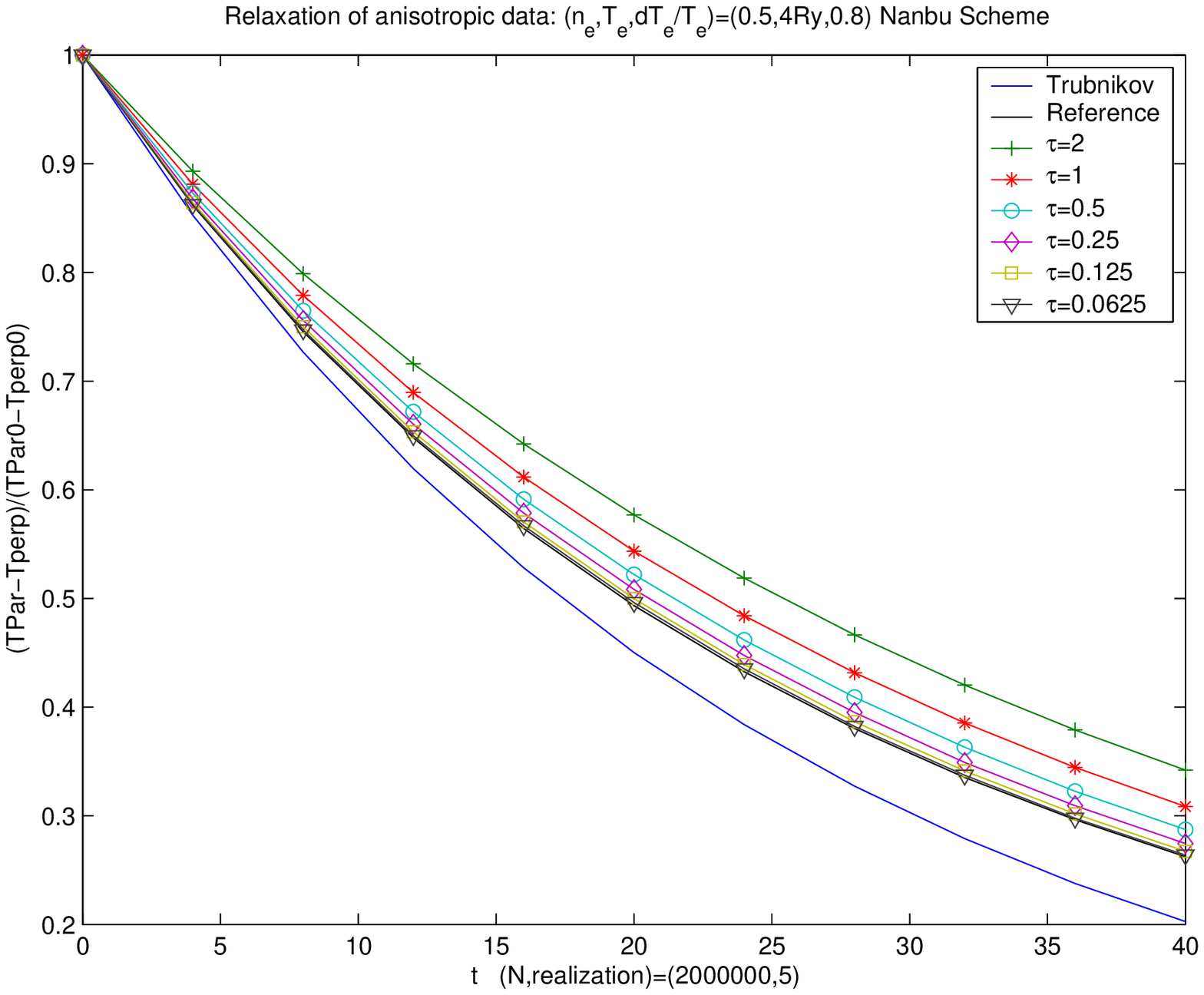}
\caption{Relaxation of the velocity distribution function with
anisotropic initial data for different values of $\tau$.
Bobylev-Nanbu scheme.}\label{fig:DSMC3}
\end{center}
\end{figure}

\begin{figure}
\begin{center}
\includegraphics[scale=0.58]{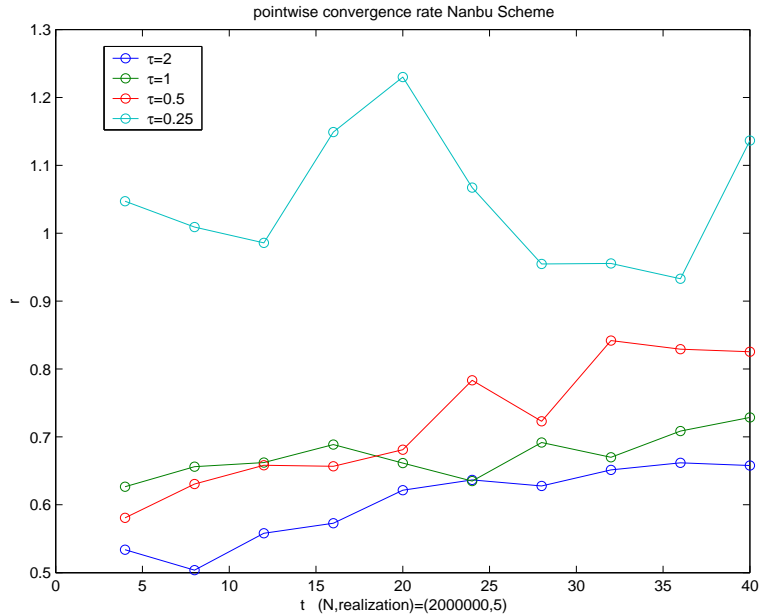}
\caption{Pointwise order of accuracy $r(\tau)=\log2(R(\tau))$ for
the Bobylev-Nanbu scheme for decreasing values of the parameter
$\tau$.}\label{fig:DSMC4}
\end{center}
\end{figure}

\begin{figure}
\begin{center}
\includegraphics[scale=0.35]{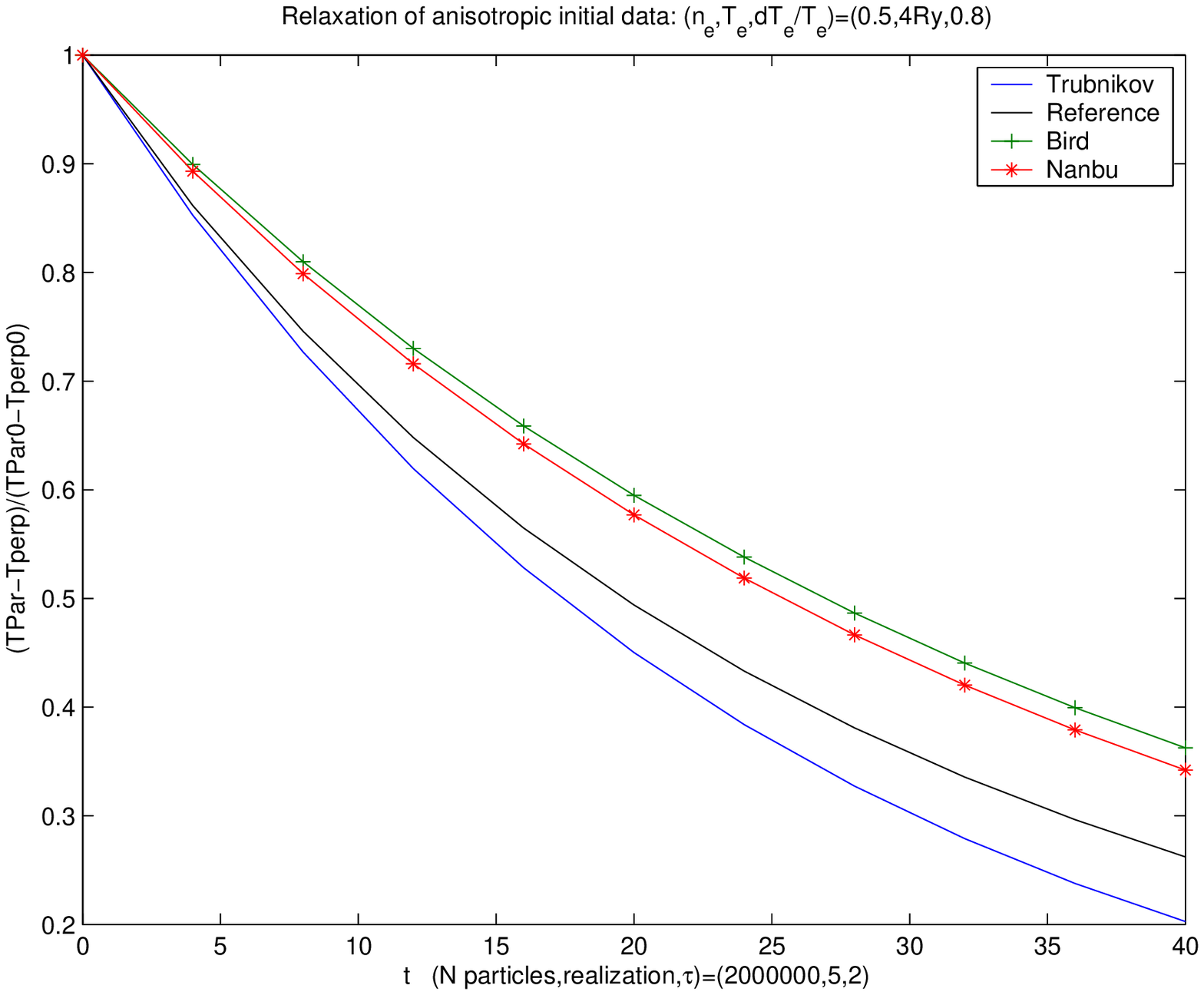}
\includegraphics[scale=0.35]{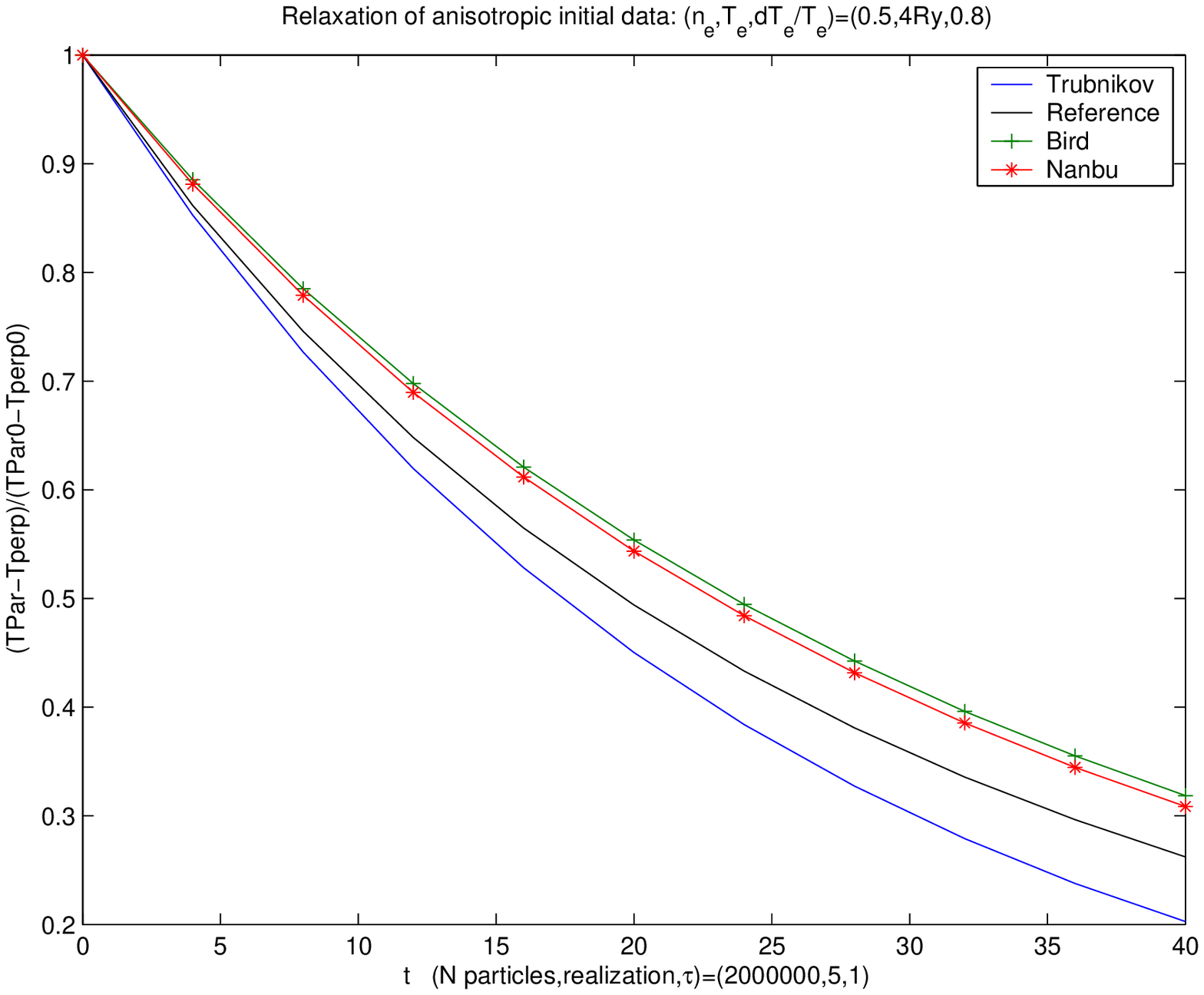}\\
\includegraphics[scale=0.35]{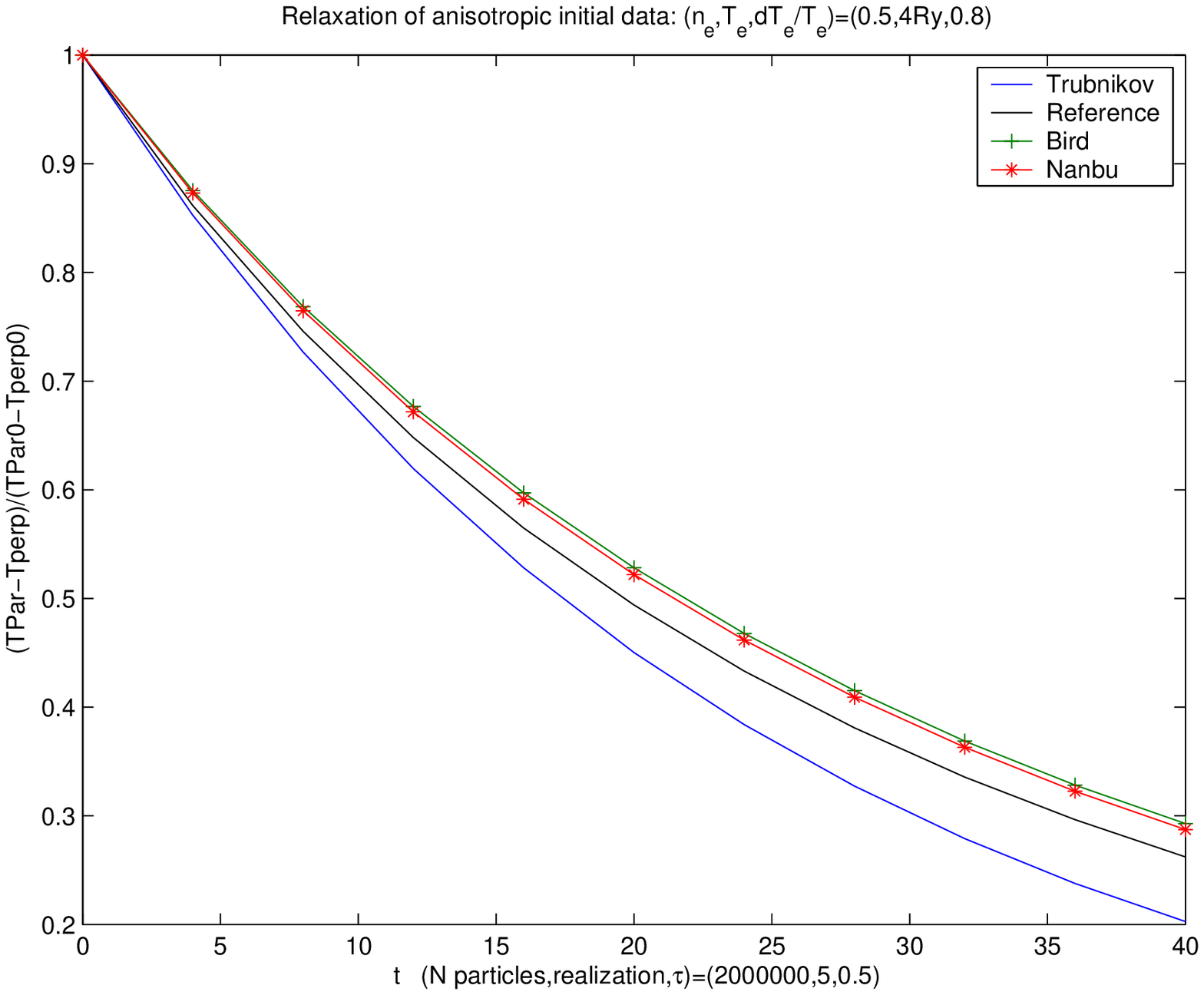}
\includegraphics[scale=0.35]{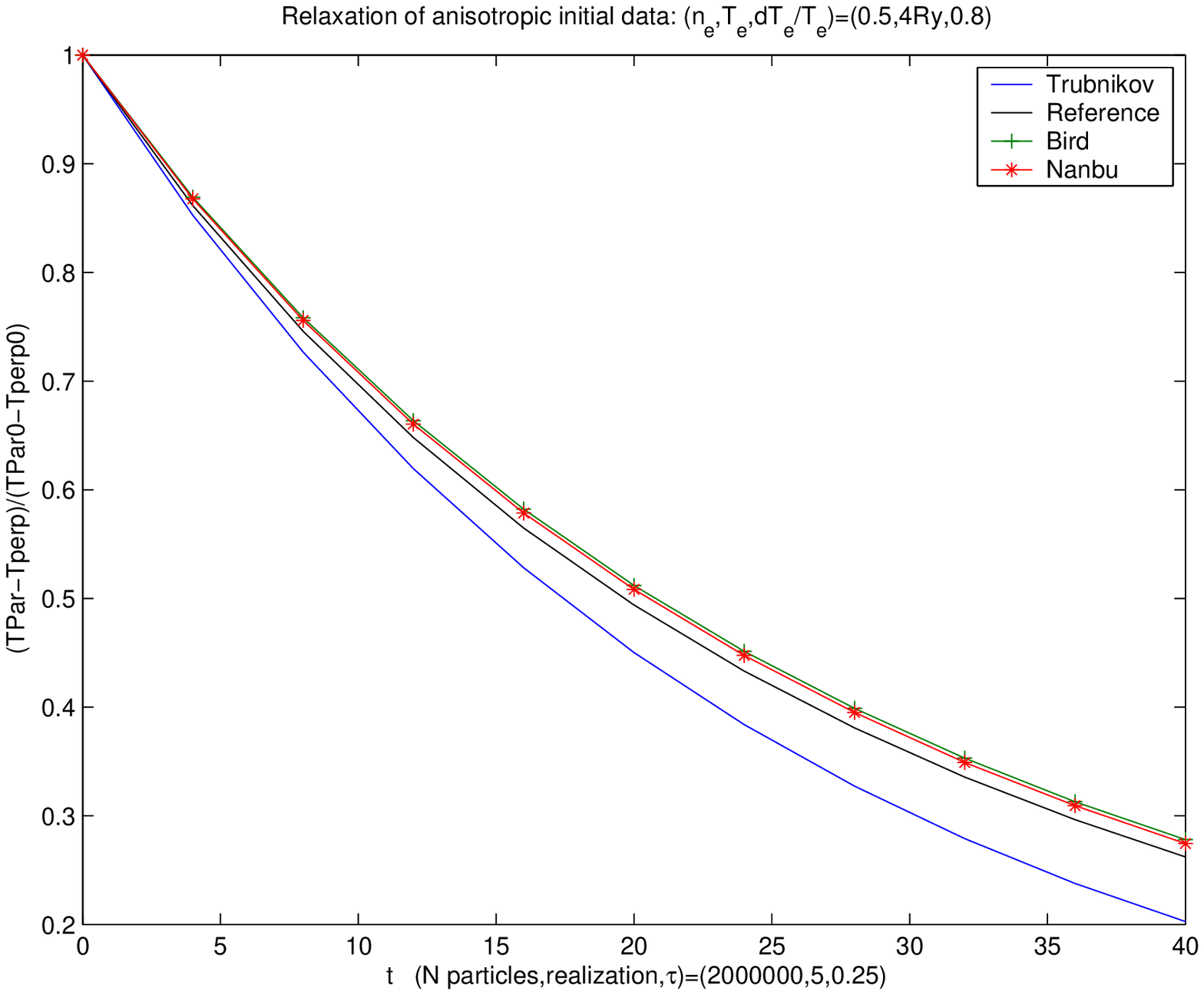}
\caption{Comparison of the Bird and Bobylev-Nanbu method for
$\tau=2$ top left, $\tau=1$ top right, $\tau=0.5$ bottom left,
$\tau=0.25$ bottom right. In each Figure the Trubnikov (blue line)
and the reference solution (black line) are
depicted.}\label{fig:DSMC5}
\end{center}
\end{figure}

Here we report the solution of the tests described in the previous
Section. In Figure \ref{fig:DSMC1} the solution for the relaxation
of temperature in the different directions is showed for the Bird
method, while in Figure \ref{fig:DSMC3} for the BN method. The
behavior of the schemes has been analyzed using six different values
for the parameter $\tau$ \be \tau=2 \ \tau=1 \ \tau=0.5 \ \tau=0.25
\ \tau=0.125 \ \tau=0.0625 \ee Moreover the solution with
$\tau=0.03125$ with $N=2\times 10^6$ and $M=10$ realizations has
been computed as a reference solution. In both Figures the analytic
and approximated solution is reported (blue line) showing a
discrepancy with the computed solution even with the more accurate
one. The time step used in the BN scheme is chosen in order to
satisfy the relation $\varrho\Delta t/\tau=1$. In Figures
\ref{fig:DSMC2} and \ref{fig:DSMC4} the convergence rate $r$ is
plotted for respectively Bird and BN. Both the schemes approach the
value $1$ when $\tau\rightarrow 0$ as expected from the theory.
Anyway it is possible to observe from Figure \ref{fig:DSMC5}, in
which the solution of the two methods for the same values of $\tau$
(respectively $\tau=2$, $\tau=1$, $\tau=0.5$ and $\tau=0.25$) has
been compared, that the two algorithm furnish a different relaxation
rate for large values of the approximation parameter of the
collision operator, while for small values the two methods in
practice coincide. This behavior can be explained observing that
while $\tau\rightarrow 0$ also the time step $\Delta t \rightarrow
0$ thus the error introduced by the BN scheme in the discretization
of the time derivative disappears.

\begin{figure}
\begin{center}
\includegraphics[scale=0.348]{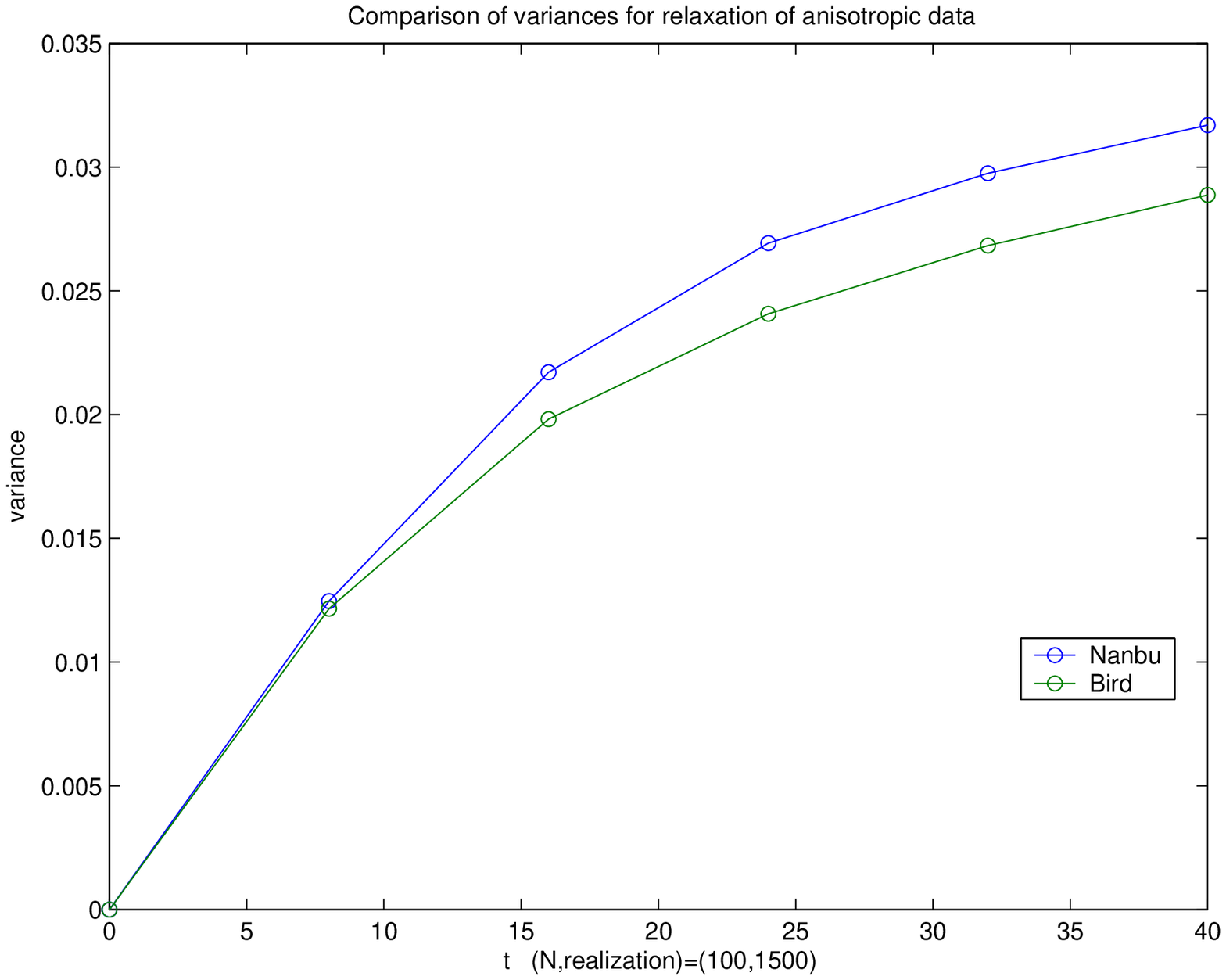}
\includegraphics[scale=0.348]{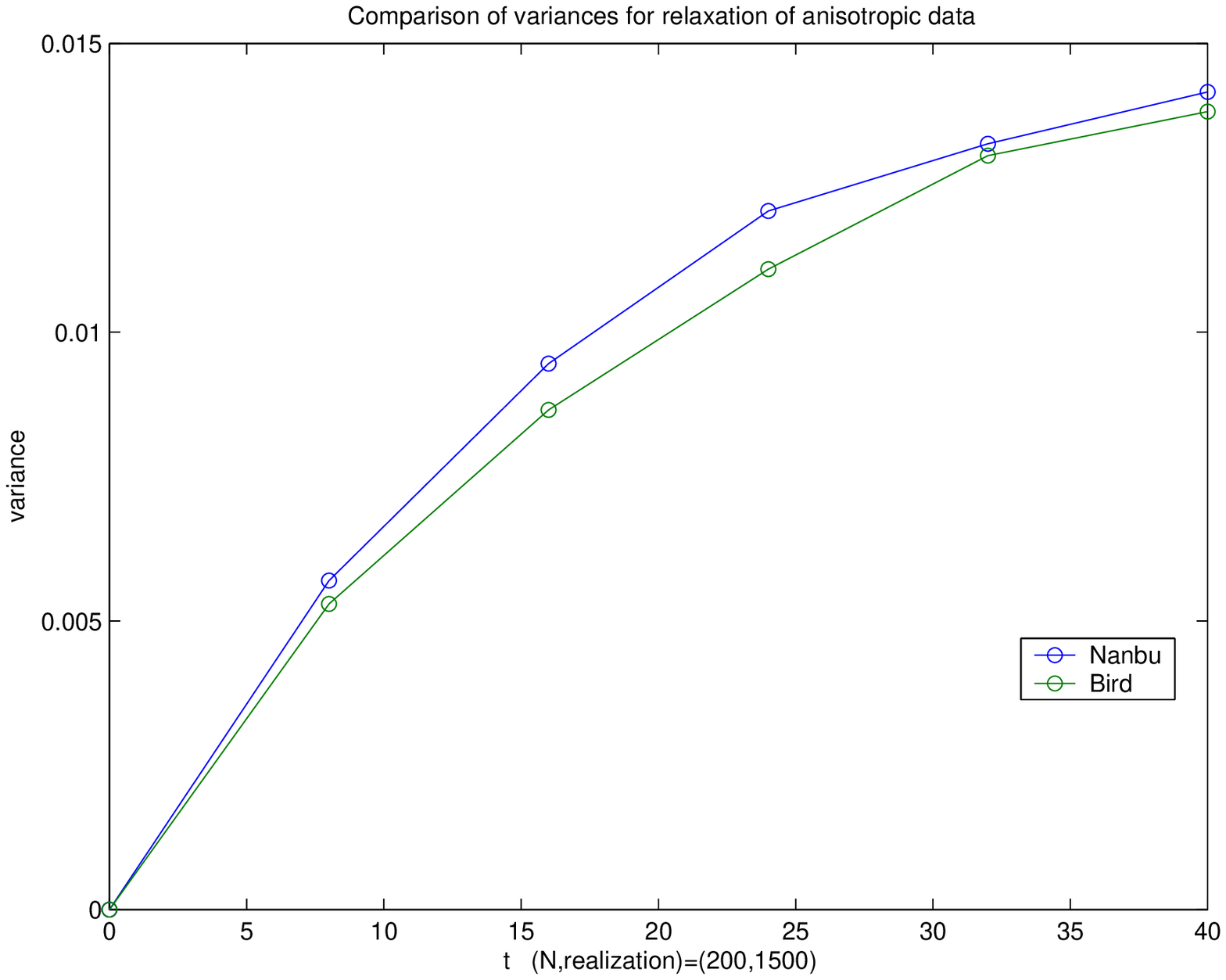}\\
\includegraphics[scale=0.35]{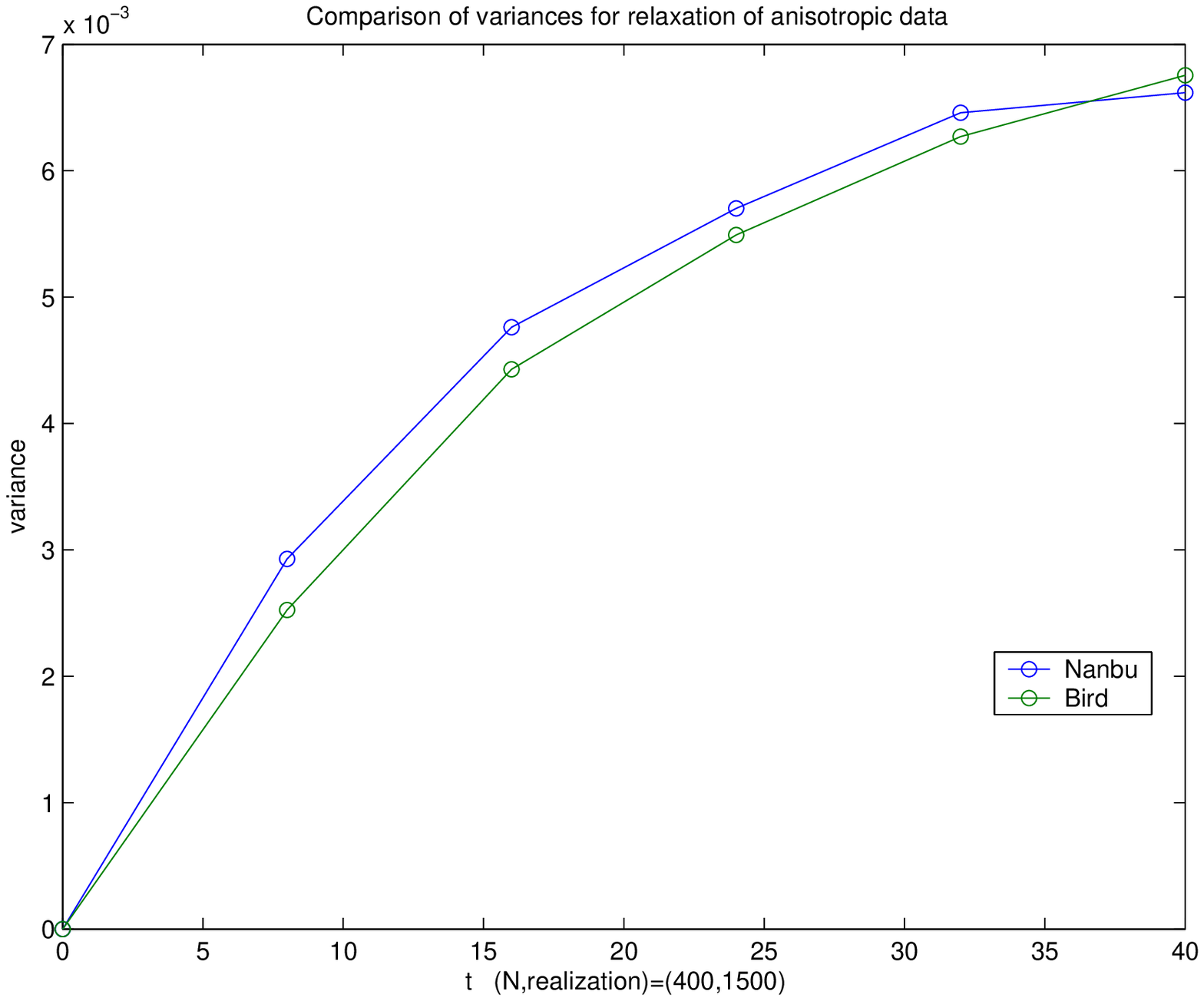}
\includegraphics[scale=0.35]{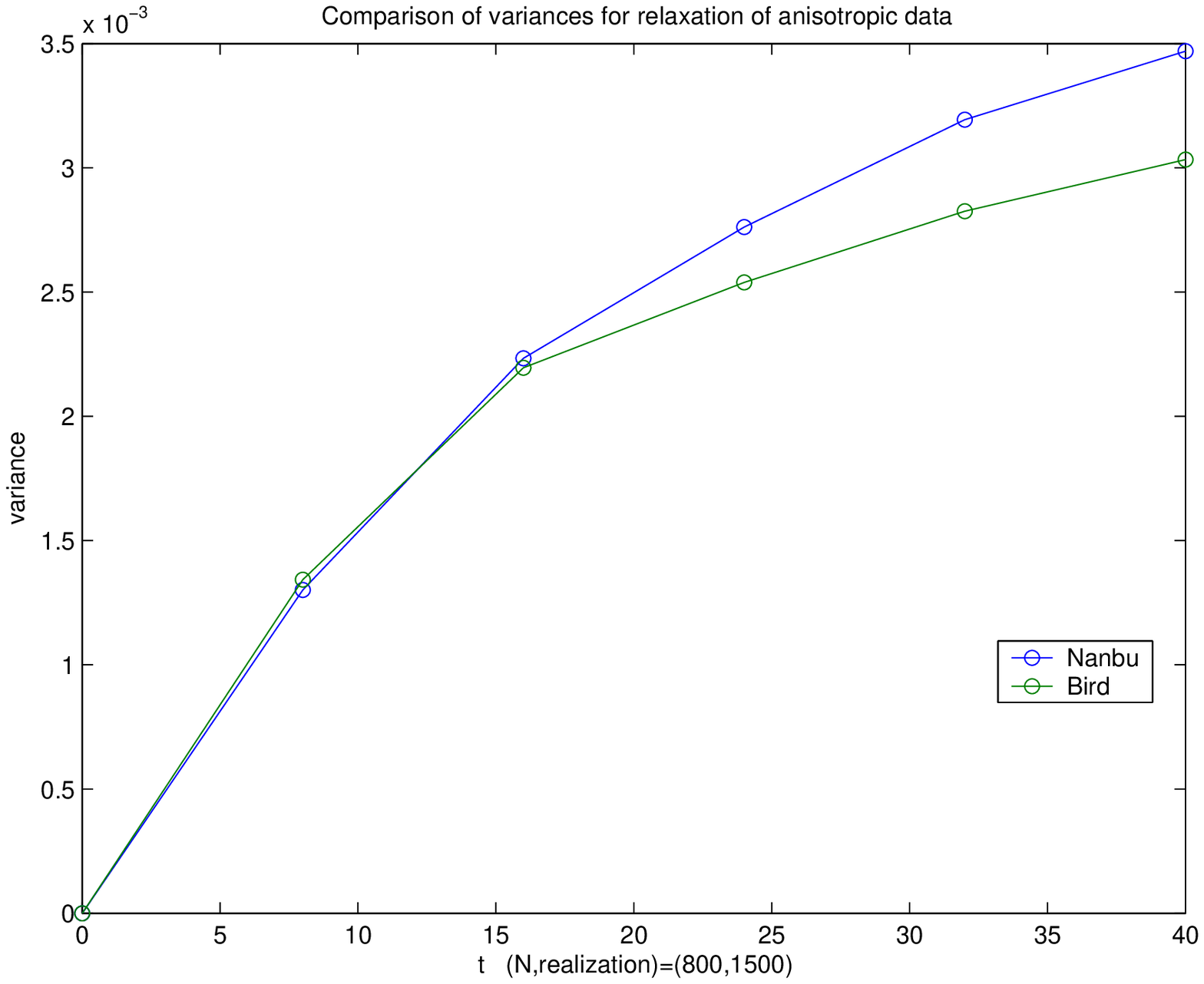}\\
\includegraphics[scale=0.35]{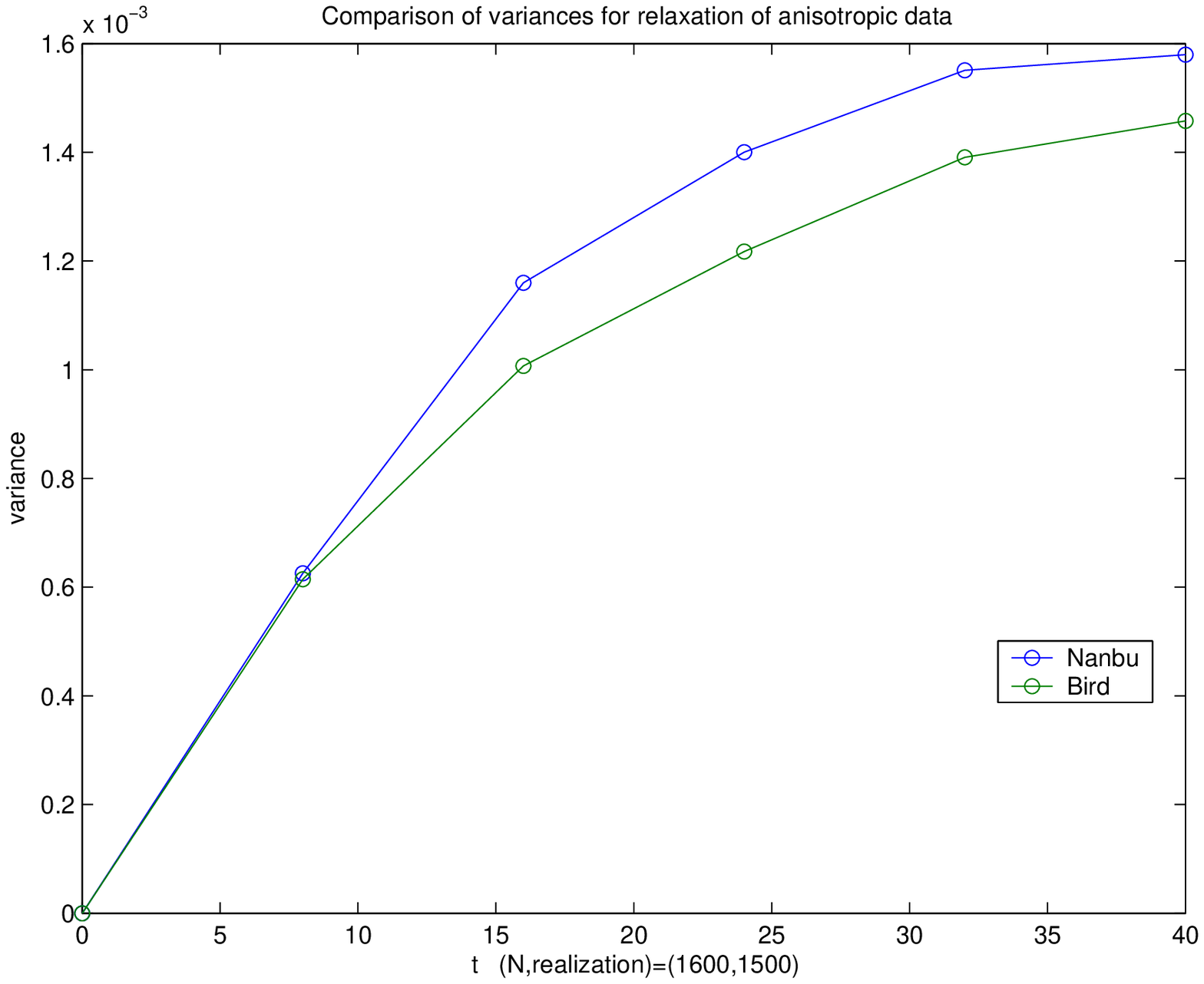}
\includegraphics[scale=0.35]{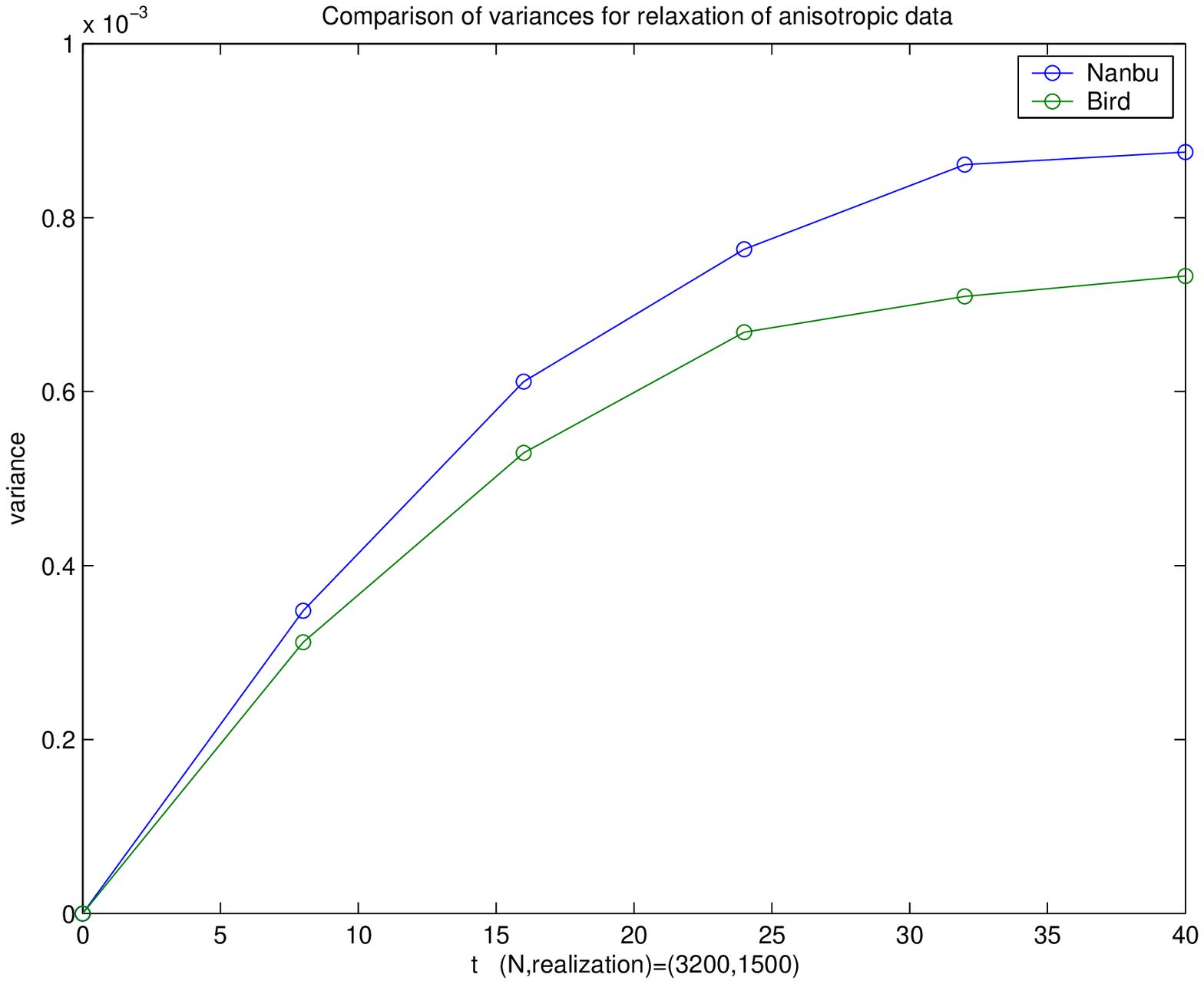}
\caption{Comparison of the Bird and Bobylev-Nanbu variance for
$\tau=1$ and $N=100$ top left $N=200$ top right $N=400$ middle left
$N=800$ middle right $N=1600$ bottom left $N=3200$ bottom right with
$M=1500$ realizations.}\label{fig:DSMC6}
\end{center}
\end{figure}

In order to show the different performance in terms of statistical
fluctuations we fixed the parameter $\tau=1$ and perform several
simulations increasing the number of samples $N$. In Figure
\ref{fig:DSMC6} a comparison between the two variances, obtained
with Bird and BN, has been plotted. The statistical fluctuations of
the two method are approximately the same for all the initial
choices of $N$; nevertheless it is possible to see how the Bird
scheme oscillates slightly less then BN scheme for all the values.
This behavior is mainly due to the presence of the time
discretization error in the BN scheme.
If, instead of choosing a large value for $\tau$ we keep it small,
the variance of the two methods become practically the same. In
fact, if we want a fine approximation of the collision operator with
the Bobylev-Nanbu method the time step has to be very small, which
means we are neglecting the time discretization error.

\section{Conclusion}

In this work we have proposed a generalization of the Monte Carlo
scheme proposed by Bobylev and Nanbu for the solution of plasma
physics problems in which the predominant collisions are of Coulomb
type. This result is achieved by extending the classic Nanbu and
Bird algorithms for rarefied gas dynamic to the case of plasma
physics. The new methods provide more accurate results for a fixed
$\tau$ with respect to the original BN algorithm without increasing
the computational cost. In the resulting algorithms not all
particles collide at each time step and some particles collide more
than once. From the physical point of view this result is
counterintuitive, in fact, all other computational models in
literature about Coulomb interactions are based on all particles
colliding simultaneously. In the limit of small values of $\tau$ all
methods become essentially equivalent.

In future we hope to extend the methods described in \cite{PR} for
rarefied gas dynamics to Coulomb interactions and to generalize the
hybrid techniques developed in \cite{dimarco1, dimarco2} to plasma
physics problems close to thermodynamic equilibrium. Another
interesting research direction consists in developing a more
accurate approximation in $\tau$ of the Landau operator. This would
allow the use of larger time steps in the simulations.




\bibliographystyle{plain}
\bibliography{DSMC_arxiv}

\begin{thebibliography}{10}

\bibitem{Babovsky}
H.~Babovsky.
\newblock On a simulation scheme for the {B}oltzmann equation.
\newblock {\em Math. Methods Appl. Sci.}, 8:223--233, 1986.

\bibitem{bird}
G.~A. Bird.
\newblock {\em Molecular gas dynamics and direct simulation of gas flows}.
\newblock Clarendon Press, 1994.

\bibitem{bittencourt}
J.~A. Bittencourt.
\newblock {\em Fundamentals of plasma physics}.
\newblock Pergamon Press, 1986.

\bibitem{Bobylev}
A.~V. Bobylev and K.~Nanbu.
\newblock Theory of collision algorithms for gases and plasmas based on the
  {B}oltzmann equation and the {L}andau-{F}okker-{P}lanck equation.
\newblock {\em Physical Review E}, 61(4):4576–--4586, 2000.

\bibitem{caflisch}
R.~E. Caflisch.
\newblock {M}onte {C}arlo and {Q}uasi-{M}onte {C}arlo methods.
\newblock {\em Acta Numerica}, pages 1--49, 1998.

\bibitem{CPmc}
R.~E. Caflisch and L.~Pareschi.
\newblock Implicit {M}onte {C}arlo methods for rarefied gas dynamics {I}: The
  space homogeneous case.
\newblock {\em J. Comput. Phys.}, (154):90--116, 1999.

\bibitem{CPima}
R.~E. Caflisch and L.~Pareschi.
\newblock Towards a hybrid method for rarefied gas dynamics.
\newblock {\em IMA Vol. App. Math.}, (135):57--73, 2004.

\bibitem{cohen}
R.~E. Caflisch, C.~Wang, G.~Dimarco, B.~Cohen, and A.~Dimits.
\newblock A hybrid method for accelerated simulation of {C}oulomb collisions in
  a plasma.
\newblock {\em Multiscale Model. Simul.}, (7):865--887, 2008.

\bibitem{cercignani}
C.~Cercignani.
\newblock {\em The {B}oltzmann Equation and Its Applications}.
\newblock Springer-Verlag, 1988.

\bibitem{Desvilletts}
L.~Desvilletts.
\newblock On asymptotics of the {B}oltzmann equation when the collisions become
  grazing.
\newblock {\em Transp. Th. Stat. Phys.}, (21):259--276, 2002.

\bibitem{dimarco1}
G.~Dimarco and L.~Pareschi.
\newblock Hybrid multiscale methods {I}. {H}yperbolic relaxation problems.
\newblock {\em Comm. Math. Sci.}, (1):87--133, 2006.

\bibitem{dimarco2}
G.~Dimarco and L.~Pareschi.
\newblock A fluid solver independent hybrid method for multiscale kinetic
  equations.
\newblock {\em SIAM J. Sci. Comp.}, to appear.

\bibitem{Candal}
A.M. Dimits, C.~Wang, R.~Caflisch, B.I. Cohen, and Y.~Huang.
\newblock Understanding the accuracy of {N}anbu's numerical {C}oulomb collision
  operator.
\newblock {\em J. Comput. Phys.}, 228:4881--4892, 2009.

\bibitem{Nanbu80}
K.~Nanbu.
\newblock Direct simulation scheme derived from the {B}oltzmann equation.
\newblock {\em J. Phys. Soc. Japan}, 49:2042--2049, 1980.

\bibitem{Nanbu}
K.~Nanbu.
\newblock Theory of cumulative small-angle collision in plasmas.
\newblock {\em Physical Review E}, 55(4):4642--4652, 1997.

\bibitem{PR}
L.~Pareschi and G.~Russo.
\newblock Time relaxed {M}onte {C}arlo methods for the {B}oltzmann equation.
\newblock {\em SIAM J. Sci. Comput.}, 23:1253--1273, 2001.

\bibitem{Abe}
T.~Takizuka and H.~Abe.
\newblock A binary collision model for plasma simulation with a particle code.
\newblock {\em J. Comp. Phys.}, 25:205--219, 1977.

\bibitem{Trubnikov}
B.~A. Trubnikov.
\newblock {\em Review of Plasma Physics}.
\newblock Consultant Bureau, 1965.

\bibitem{wang}
C.~Wang, T.~Lin, R.~E. Caflisch, B.~Cohen, and A.~Dimits.
\newblock Particle simulation of {C}oulomb collisions: Comparing the methods of
  {T}akizuka \& {A}be and {N}anbu.
\newblock {\em J. Comp. Phys.}, 227:4308--4329, 2008.

\bibitem{wang1}
W.~X. Wang, M.~Okamoto, N.~Nakajima, and S.~Murakami.
\newblock Vector implementation of nonlinear {M}onte {C}arlo {C}oulomb
  collisions.
\newblock {\em J. Comp. Phys.}, 128:209--222, 1996.

\end{thebibliography}

\end{document}